\documentclass{article}

\usepackage{PRIMEarxiv}

\usepackage[utf8]{inputenc} 
\usepackage[T1]{fontenc}    
\usepackage{hyperref}       
\usepackage{url}            
\usepackage{booktabs}       
\usepackage{amsfonts}       
\usepackage{nicefrac}       
\usepackage{microtype}      
\usepackage{lipsum}
\usepackage{fancyhdr}       
\usepackage{graphicx}       
\graphicspath{{media/}}     
\usepackage{float}
\usepackage{amsmath}
\usepackage{multirow}

\title{Analysis of h-index for research awards
}

\author{
  Aashay Singhal \\
  IIIT Hyderabad \\
  \texttt{aashay.singhal@research.iiit.ac.in} \\
   \And
  Kamalakar Karlapalem \\
  IIIT Hyderabad \\
  \texttt{kamal@iiit.ac.in} \\
}

\begin{document}
\maketitle



\section{Abstract}
In order to advance academic research, it is important to assess and
evaluate the academic influence of researchers and the findings they
produce. Citation metrics are universally used methods to evaluate
researchers. Amongst the several variations of citation metrics,
the h-index proposed by Hirsch has become the leading measure.
Recent work shows that h-index is not an effective measure to
determine scientific impact - due to changing authorship patterns.
This can be mitigated by using h-index of a paper to compute h-
index of an author. We show that using fractional allocation of
h-index gives better results. In this work, we reapply two indices
based on the h-index of a single paper. The indices are referred to as:
hp-index and hp-frac-index. We run large-scale experiments in three
different fields with about a million publications and 3,000 authors. We also compare h-index of a paper with nine h-index like metrics.
Our experiments show that hp-frac-index provides a unique ranking
when compared to h-index. It also performs better than h-index in
providing higher ranks to the awarded researcher.

\section{Introduction}
In the world of academia, the h-index has become a popular measure of an author's research impact. Introduced by Jorge Hirsch in 2005, the h-index reflects both the productivity and impact of a researcher's work by taking into account the number of publications and the number of citations they have received. The h-index is calculated by determining the number of articles published by a researcher and the number of citations each article has received. A researcher has an h-index of $h$ if $h$ of their articles have been cited at least $h$ times. The h-index is a useful tool for both researchers and academic institutions because it provides a quantitative measure of a researcher's scholarly output and impact. The h-index is particularly valuable when comparing researchers from different fields or when evaluating a researcher's entire career, as it accounts for both the number of publications and the quality of those publications.

    The h-index has become a popular tool for evaluating the research impact of scholars because it is relatively simple to calculate and provides a single number that can be compared across researchers. However, the h-index has some limitations, and it should not be the sole measure of a researcher's productivity or impact. For example, the h-index does not take into account the context of citations, such as whether they are from highly respected journals or from less reputable sources. Additionally, the h-index can be affected by factors outside of a researcher's control, such as the size of their research community or the time period in which their work was published. Additionally, the h-index is heavily influenced by a researcher's most highly cited papers, which may not necessarily reflect the overall impact of their work. The h-index is discipline-specific and does not account for the varying citation practices across different fields, which may result in unfair comparisons between researchers working in different disciplines. Nevertheless, despite its limitations, the h-index remains a widely used metric in academia for evaluating the research output and impact of scholars.

There are several variations of the h-index that have been proposed to address its limitations. One such variation is the g-index \cite{egghe2006theory}, which takes into account the number of highly cited papers a researcher has published. Another variation is the m-index \cite{bornmann2008there}, which measures the productivity of a researcher by dividing the h-index by the number of years since their first publication. Other variations include the a-index \cite{jin2007r}, which measures the author's total number of citations. These variations provide more nuanced insights into a researcher's impact and productivity, but they also have their own limitations and should be used in conjunction with other metrics.

The h-index is typically used to evaluate the overall productivity and impact of a researcher's work, but it is also possible to calculate the h-index for individual publications. This can be a useful tool for evaluating the impact of a particular article or book within a field. To calculate the h-index of a single publication, you would need to determine how many times that publication has been cited, and then identify the number $n$ of other publications that have been cited at least $n$ times. If the publication in question has an h-index of $n$, this means that it has been cited at least $n$ times, and there are at least $n$ other publications that have been cited at least $n$ times. Calculating the h-index of single publications can be useful for researchers who want to understand the impact of their work, or for publishers and editors who want to evaluate the quality of submissions.

In order to understand whether h-index is a good evaluation metric for single publications, we compared it with nine variations of h-index. We collected the papers for VLDB and SIGMOD conference over the years and necessary data required (like their citations, year of publishing, etc.) to calculate the metrics on each of the paper. After calculating each metric value on each paper, we rank them by the metrics. We also gather the awarded papers in each of the two conferences. After comparing ranks of the awardees, our results show that h-index is the best in ranking awardees at the top. Among our experiments, we also discuss the correlation and overlap of these metrics.

After defining the h-index of a paper, the next natural step is to use this h-index to compute the impact of a researcher. Here, we apply four h-index like metrics to evaluate researchers, namely, h-index, h-frac-index, hp-index and hp-frac-index. Amongst these four metrics three of them have been proposed in prior works \cite{hirsch2005index, koltun2021h, egghe2011single} but
hp-frac-index is a newly proposed metric in our work. The traditional h-index uses number of citations of the papers published by an author to calculate the h-index of an author. In hp-frac-index, instead of number of citations we use h-index of the published papers to calculate h-index of an author. Further, we divide the h-index of each paper by the number of authors. 

We collect top 1000 researchers in the field of Computer Science, Economics and Biology. Then we gather all the papers published by them and other necessary data required to calculate each of the four metrics. We also gather the list of awarded researchers amongst the list we obtained. Then we calculate the four metrics on each researcher and rank them. On comparing the ranks given by each metric, we find that hp-frac-index is better than the other metrics in ranking the awardees at top. The hp-frac-index is a reliable means of assessing the influence of researchers. hp-frac-index is resistant to manipulation and can capture individual contributions effectively. These combined factors make the hp-frac-index superior in ranking authors.

\noindent \textbf{Contributions:}

\begin{itemize}
    \item We define h-index of a paper and determine which variation of h-index like metrics ranks the awarded papers higher
    \item We use h-index of a paper to come up with hp-frac-index applied for author impact. We then compare hp-frac-index with other h-index like indices to analyse their performance in ranking awarded researchers.
\end{itemize}

\section{H-index and its variants on research papers}

Finding the most relevant scientific article from a set of articles may seem to be a simple task at first sight, but the task to rank the articles is specially challenging. Impact of a publication is one of the most important topics in scientometrics.
The sheer increase in number of publications per year has made it hard for researchers to keep track of the literature. This problem of inflation in scientific articles makes it a challenging task to find papers that have made significant contributions. This is especially true for the newcomers in the field. 

The evaluation of a single publication serves as the foundation for evaluating scientists, organisations, journals, and other aspects of scientific research outputs. Today, citation counts is the most widely used quantitative method to evaluate single publication. However, citation counts can only roughly reflect a publication's impact. Moreover, it
cannot effectively reflect a publication’s comprehensive influence (i.e. influence beyond just the first level of citations). In recent years, another source of evaluation has emerged which measures the impact of an article in society: alternative metrics (atlmetrics \cite{priem2011altmetrics}). Altmetrics are alternative approaches to measuring the impact of a research article, as demonstrated by users' interest and engagement with it. Altmetric watches social media sites, science blogs, many mainstream media outlets and reference managers for mentions of academic papers. Some of the metrics are as follows: number of views, downloads, clicks, saves, tweets, shares, posts, discussions, and bookmarks. These altmetrics are available on SCOPUS \cite{scopus} and PLOS \cite{plos}. Altmetrics aim to complement traditional research impact measures by showing a more complete picture of how readers engage with and use it.

Schubert \cite{schubert2009using} proposed to use h-index for assessing single publications. Since the h-index can be used to assess individual articles, we should also use other Hirsch-type indices in assessing single publications. In the current research, eight other variants of h-index (including the original h-index) and one traditional indicator (number of papers). Therefore, nine indicators in total, were chosen to be used for assessing individual articles.

 \hfill \break
\noindent\textbf{Contributions:}
In this section, we answer the following questions:
\begin{itemize}

\item Determine which variations of the h-index have most overlap with each other. For discussing this, we calculate Rank Biased Overlap (RBO).
\item Which variant gives the best ranking to national and international award winning authors? 
\item Does the performance of these indices change over time? 
\end{itemize}

\subsection{Data}
In order to evaluate different citation based indices, data for a large scale of papers is required. It is important to collect suitable data in order to answer the research questions in this work. So we have selected papers from VLDB conference and SIGMOD conference. There are multiple reasons for the selection of these specific conferences. Firstly, their list of papers is easily available on DBLP (dblp.org). Both these conferences are amongst the top conferences in their fields. They have dedicated committees who select the test of time awards each year which we will use later in this section. Also, the details for the same are widely available in public domain \cite{vldbtot,sigmodtot}
\subsubsection{Data Collection}
For the purposes of this research, we collect all the papers in both the conferences over the years. For calculating the indices, we also collect the citation data for these papers. 
    
For collecting these papers and citation data, we implement a three step crawler. The crawler works as follows:
\begin{itemize}
    \item it queries dblp to retreive all the paper titles and their year of publishing. This is written using beautifulSoup library in python.
    \item using S2AG API \cite{s2ag}, we map these paper titles to the paper IDs in semantic scholar. We have used semantic scholar as it has a easily available API for programmatic retrieval of data. Moreover, it has a vast coverage of papers and citation data. This retrieval process is semi-automatic. Firstly, the script searches for papers with similar title as the given title (from dblp). It then filters papers with the same publishing year. Lastly, it tries to do a fuzzy string match on paper title and if the confidence is very high, we assign this paper ID to the given title. In case the confidence is low, the user is prompted with a question and has to manually select if the two papers match. This last step is the only part where we need human intervention. Out of 8070 (4427+3643; see Table \ref{tab:dataset_desc}) papers, we only required the last step for 188 papers. 
    \item once we have the semantic scholar paper IDs (s2\_id), we can retrieve the citations, year, and other metadata. Using the S2AG API, we retrieve papers that directly cite the collected papers from above. We call this set of papers the 1th Generation publications. Then, we also retrieve papers that directly cite the 1th Generation publications. 
\end{itemize}
At the end of this process, we get a citation graph with the following sets of nodes: (a) all papers from both the conferences (b) their corresponding citations and (c) citations of these citations. We also have the year of publishing for all of these papers. The edges in the graph denote the 'is cited by' relation. In other words, an edge from node A to node B denotes that A is cited by B. 

\subsubsection{Dataset Description}
We collected the papers from DBLP for the years 1985-2020 for VLDB and 1988-2020 for SIGMOD. As shown in table \ref{tab:dataset_desc} below, there were 4652 and 3744 papers listed on DBLP respectively. Out of which we were able to map more than $95\%$ of papers to their corresponding semantic scholar paper ID. This was done using our fuzzy match logic on paper titles. After this, we retreived 174669 direct citations for the 4427 papers from VLDB and 177160 for 3744 papers from SIGMOD. For the next level, we retreived 1312443 and 1378127 citations of the 1th generation papers. 0th generation to 1th generation is a 100x increase in the number of papers but 1th generation to 2th generation is a 10x increase. Finally, we see total papers in the citation graph as 1316634 and 1382516 respectively.

\begin{table}[H]
    \centering
    \begin{tabular}{l c c }
    \hline
         & VLDB & SIGMOD  \\
    \hline
        DBLP Papers & 4652 & 3744 \\
        Papers with $s2\_id$ & 4427 & 3643 \\
        1th gen papers & 174669 & 177160 \\
        2th gen papers & 1312443  & 1378127 \\
        
        Total papers (nodes) & 1316634  & 1382516 \\
        Total cites relation (edges) & 3135134 & 3254607   \\
        
    \hline
    \end{tabular}
    \caption{Dataset Description}
    \label{tab:dataset_desc}
\end{table}

\subsubsection{Benchmark Dataset}
This specific research problem has no gold standard based on a dataset that could be used to evaluate and assess. A comprehensive and extensive benchmark dataset is required to assess the indices. Hence, in this study, the test of time awards are used as a standard merit or benchmark. In the context of VLDB, a paper is selected from the VLDB Conference from ten to twelve years earlier that best meets the “test of time”. In picking a winner, the committee evaluates the impact of the paper. The committee especially values impact of the paper in practice, e.g., in products and services. Impact on the academic community demonstrated through significant follow-through research by the community is also valued. For SIGMOD, this paper is selected from the conference held exactly ten years ago. Their criterion of identifying the paper is impact (research, products, methodology) over the intervening decade.

\begin{table}
    \centering
    \begin{tabular}{l c c }
    \hline
         & VLDB & SIGMOD  \\
    \hline
        Total awarded papers & 29 & 25 \\
        Total awarded papers in our dataset & 25 & 21 \\
    \hline
    \end{tabular}
    \caption{Number of awarded papers}
    \label{tab:benchdataset_desc}
\end{table}

In this study, we retrieve all the awarded papers for both the conferences. This data is available on their respective websites. The dataset consists of awardees from 1995 to 2022 for VLDB conference and from 1999 to 2022 for SIGMOD. Total awardees are listed in the table \ref{tab:benchdataset_desc}. The awardees for a few years are not present as either there was no award in a particular year or the corresponding paper did not exist in the crawled dataset. The reason for it missing from the dataset is that it is missing in the semantic scholar database. There are 4 out of 25 missing in SIGMOD dataset and 4 out of 29 missing in VLDB.

\subsection{Experiments}
In this section, we explain three experiments conducted in order to answer the research questions in this work. Firstly, the correlations are evaluated between the h-index and all its variants. We have also evaluated whether the awarded papers rank on the top by using h-index and its variants. We used the test of time awards that are won by papers for their exceptional impact and performance in a decade to serve as a benchmark (details in the section above). Lastly, we have compared the performance of h-index and its variants by considering change through time.

\subsubsection{Rank Biased Overlap (RBO)}
In this experiment, we compare the different variations of the h-index on the basis of their overlap with each other in ranking the papers. Unlike correlation measures, RBO is a similarity measure which denotes how similar are two ranked lists. RBO is based on the simple concept of average set overlap. The idea is to determine the fraction of content overlapping at different depths in the two ranked lists. Suppose we have two ranked lists, $A: [P_1,P_2,P_3,P_4]$ and $B: [P_2,P_1,P_4,P_3]$ in order of their ranks. Given below are set intersections at different depths. Set intersection shows the intersection between the two sets of lists at each depth. Fraction denotes the length of intersection set divided by the depth.
\begin{table}[h]
    \centering
    \begin{tabular}{l l l l c}
    \hline
        Depth(d) & Items in List A@d & Items in List B@d & set intersection & Fraction  \\
    \hline
       1 & $P_1$ & $P_2$ & $\{\}$ & 0/1=0 \\
    2 &  $P_1,P_2$ & $P_2,P_1$ & \{$P_1$, $P_2$\} & 2/2=1 \\
       3& $P_1,P_2,P_3$ & $P_2,P_1,P_4$ & \{$P_1$, $P_2$\} & 2/3=0.66 \\
    4&    $P_1,P_2,P_3,P_4$ & $P_2,P_1,P_4,P_3$ & \{$P_1$, $P_2$, $P_3$, $P_4$\} & 4/4=1 \\
        
    \hline
    \end{tabular}
    \caption{Example of set overlap calculation}
    \label{tab:rbo_eg}
\end{table}

After calculating the fractions of set overlapping at various depth, one can either plot the distribution to study how similar two lists are or, use the average of the last column (Fraction) to denote the Average overlap. RBO is a further extension of this concept which uses fixed weights for each depth. It uses a geometrically decreasing series for the weights for each depth. This makes the final value to be bound as the sum of indefinite geometric series is finite. RBO also gives higher importance to the top ranks as compared to the lower ranks due to this geometrically decreasing series. The equation for RBO is denoted by, 

\begin{equation}\label{}
    RBO(S, T, p) = (1 - p)  \sum_{d=1}^{\infty} p^{d-1} . A_d
\end{equation}
The value of RBO lies between 0 and 1 (inclusive) where 0 denotes completely disjoint ranked lists and 1 denotes identical ranked lists. 

The key difference between correlation and RBO is that the former is used to evaluate the similarity in the trends of ranking and the latter is used to evaluate the overlap in two ranked lists. RBO also gives more weightage to the top ranks in the ranked lists. In other words, a mismatch in top ranks is given more importance in the final value of RBO. 

Similar to correlation, we gather all papers for both VLDB and SIGMOD conference, then calculate all the indices for each paper. Therefore, creating nine different ranked lists. Then these ranked lists are compared pair wise. 
\subsubsection{Performance of indices in predicting trends of awardees}
In this experiment, we answer the question: Which variant is the best in ranking the awarded papers on top amongst the possible candidates. The data for the same is prepared as follows: 
\begin{itemize}
    \item gather the list of all \textit{test of time} awarded papers from VLDB and SIGMOD conferences shown in table \ref{tab:benchdataset_desc}. 
    \item iterate over all the awarded papers. For each such paper, we retrieve the set of candidates as the papers published from 10 to 12 years ago in the same conference for VLDB. For SIGMOD, the candidates are from the SIGMOD proceedings exactly 10 years ago. For example, while considering an awarded paper in SIGMOD 2020, we will pickup all the papers from SIGMOD 2010 as the candidates for this award.
    \item Using the crawled citation graph, we then retrieve the citation data for these candidates up to the year in which the award was given. This way we will get the data that the awarding committee uses while selecting. For example, while considering an awarded paper in SIGMOD 2020, we will consider the citation data only up to 2020. In other words, any citation received in 2021 will not be considered.
    \item rank all the candidates for this particular award using the nine different indices. 
    \item retrieve the rank of this particular awarded paper with respect to each index. 
\end{itemize}
Finally, after calculating these ranks for each awarded paper, we determine how many papers are present in the top 5\% of their corresponding list. We also find out the occurrence of awardees in 5–10\%, 11–20\% up to 31–40\%. 

For instance, consider an awarded paper in VLDB for year 2018 named $P_1$, we take all VLDB papers from 2006, 2007 and 2008 as the candidates for ranking. We then rank these candidates considering the citation data till 2018 only as this is the data available to the awarding committee at the time of selection. Let's say the rank for $P_1$ as per the h-index is $2$ out of $100$. Therefore, $P_1$ ranks in the top 5\% of its list according to h-index.

\subsubsection{Performance of indices over time}
In this experiment, we analyse the performance of each index over time. The measure of performance is the number of papers the index ranks in the top $5\%$ of their corresponding list. The time range is considered from 1st year after publication to 10th year after publication. For example, consider a awarded paper $P$ published in $2004$, we will evaluate the rank for this paper as per each index in the years $2005$ to $2014$. Let's say this paper in amongst the top $5\%$ in year 2007 as per h-index. We will increment the performance measure of h-index in year 2007 by one. Wherever this rank is in the top $5\%$, it will be counted towards the performance of that index in that particular year. 

\subsection{Results}

\subsubsection{Rank Biased Overlap (RBO)}
\begin{figure}[H]
    \centering
    \includegraphics[scale=0.6]{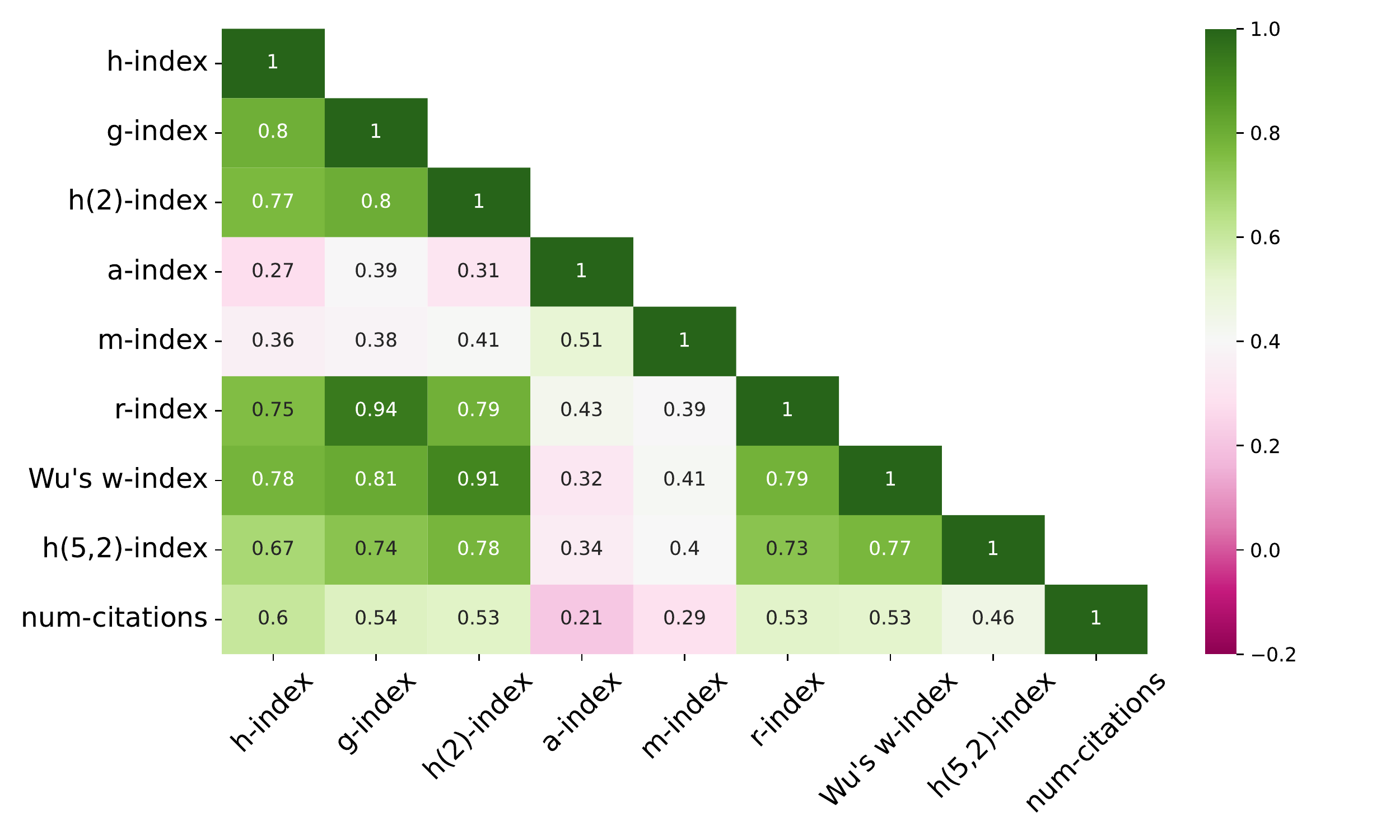}
    \caption{RBO matrix for SIGMOD conference}
    \label{fig:rbo_sigmod}
\end{figure}

\begin{figure}[H]
    \centering
    \includegraphics[scale=0.6]{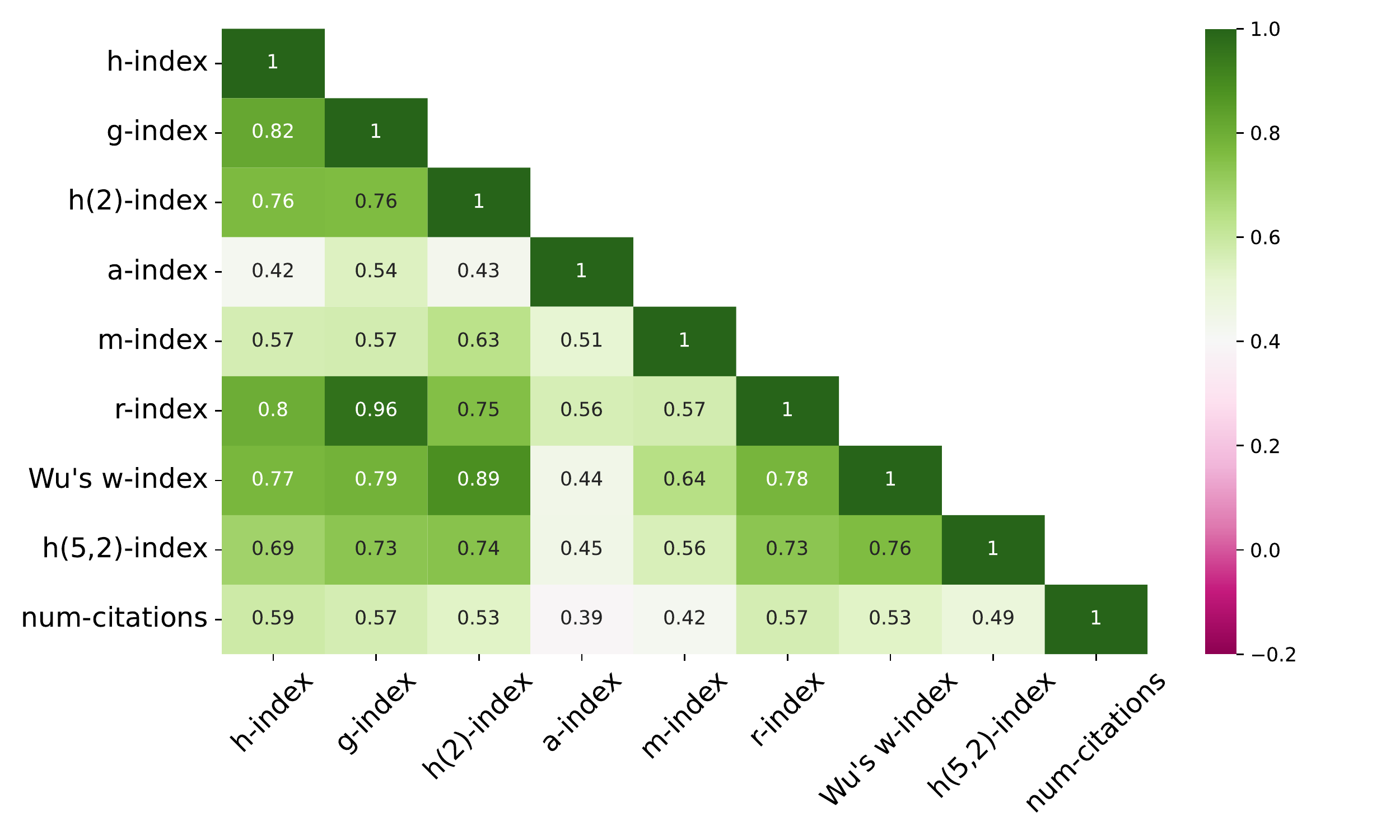}
    \caption{RBO matrix for VLDB conference}
    \label{fig:rbo_vldb}
\end{figure}

The pairwise RBO values are shown as matrix in Fig \ref{fig:rbo_sigmod} and \ref{fig:rbo_vldb}. We can see that a-index and m-index have very low overlap with every other index. They have moderate overlap of $0.51$ with each other. The a-index and m-index have similar approach which are very different from other indices. The Num of citations has the most overlap with h-index ($0.6$) amongst all indices. Also, both matrices are very similar in trends of overlap but there is difference in magnitude of overlaps. h-index has the most overlap with g-index. 

\subsubsection{Performance of indices in predicting trends of awardees}
\begin{figure}[h]
    \centering
    \includegraphics[scale=0.6]{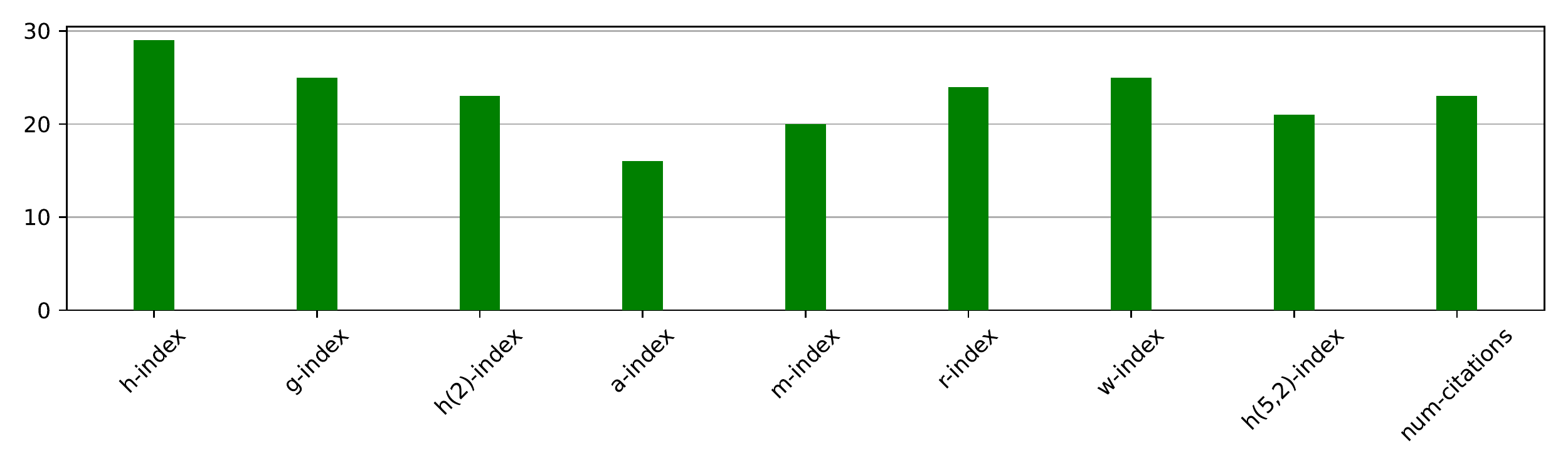}
    \caption{Index name vs Number of awarded papers ranked in top $5\%$ }
    \label{fig:exp2_1}
\end{figure}

Here we have addressed the second research question, i.e. which variants is the best in ranking the awarded papers at the top. As explained in detail in Section 5.3.2, once we have ranks of all awarded papers in their respective list for all of the indices, we first evaluate how many of the awardees were present in the top 5\% of their lists according to each index. From Fig. \ref{fig:exp2_1}, we can see that h-index performs the best with 29 out of 46 (63\%) of papers in top 5\%. The g-index, h(2)-index, r-index, w-index and number of citations show similar performance of around 52\%. The a-index is the worst performing at 34\% (16/46).

\begin{figure}[h]
    \centering
    \includegraphics[scale=0.6]{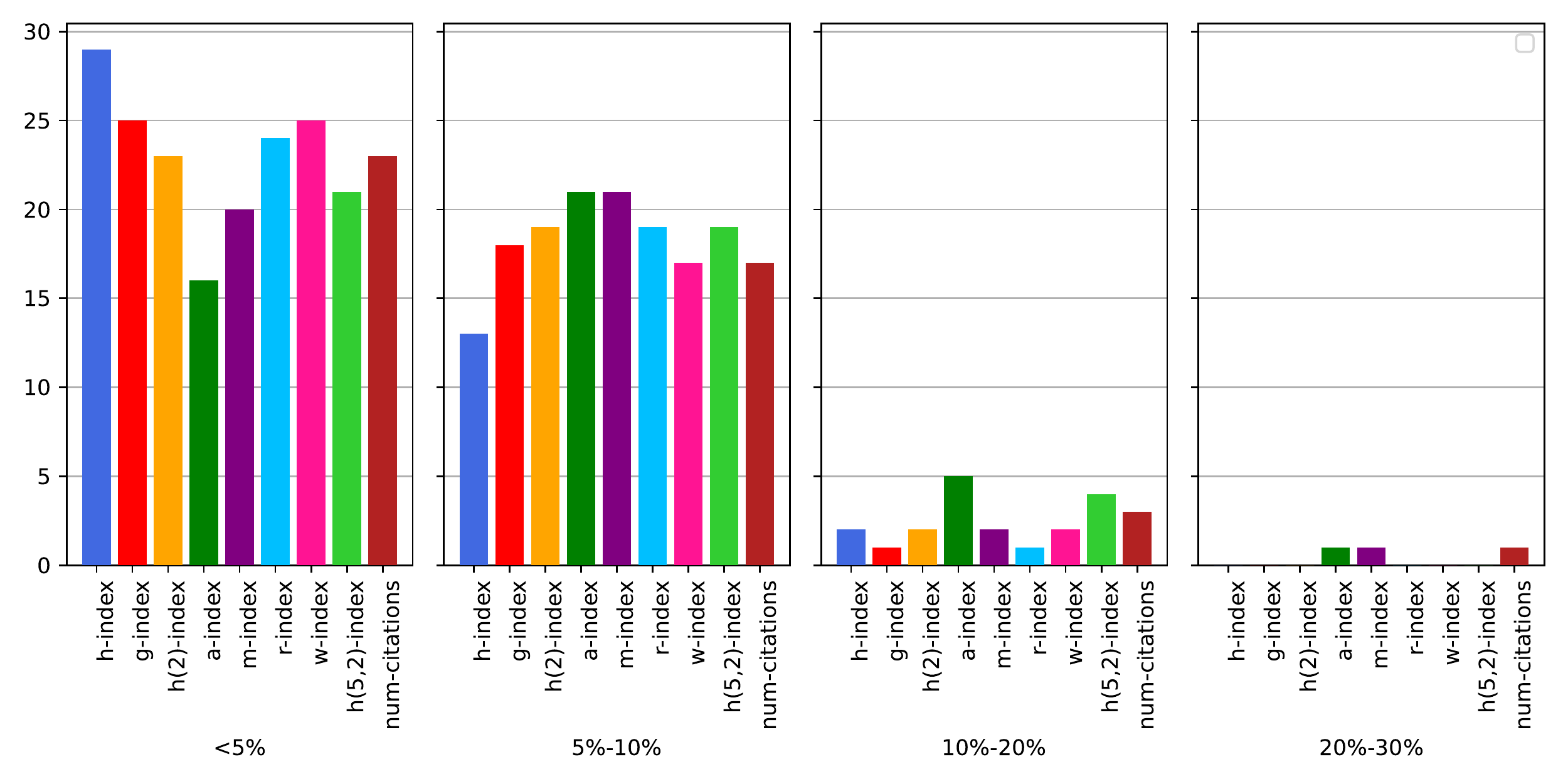}
    \caption{Index name vs Number of awarded papers ranked in $<5\%$, $5\%-10\%$, $10\%-20\%$ and $20\%-30\%$  }
    \label{fig:exp2_2}
\end{figure}

\subsubsection{Performance of indices over time}

\begin{figure}[H]
    \centering
    \includegraphics[scale=0.6]{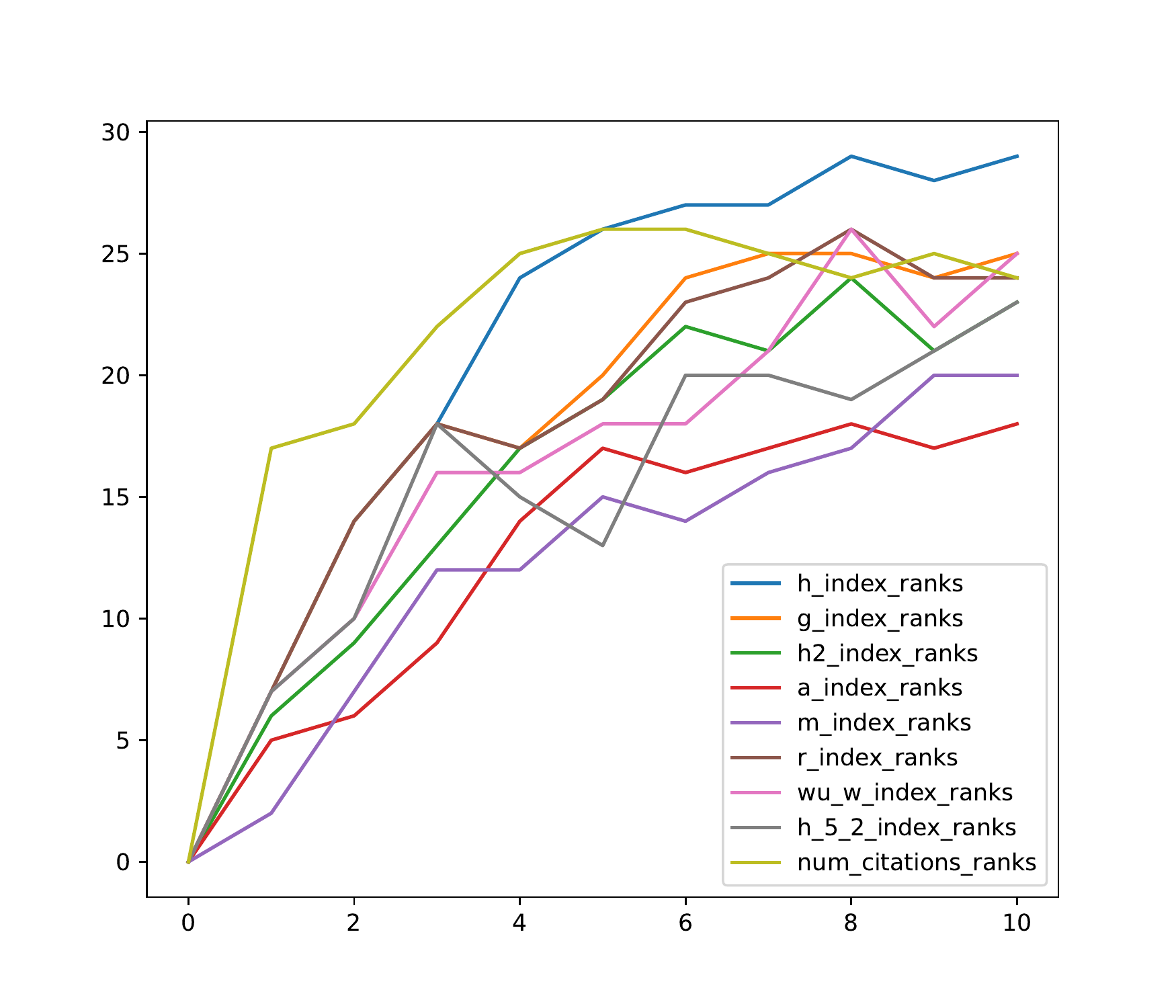}
    \caption{Number of years since publishing vs Number of papers in top $5\%$ when ranked on the particular index}
    \label{fig:time_rank}
\end{figure}

Here, we analyse the change in performance of the different indices over 10 years from publishing of the paper. From the fig. \ref{fig:time_rank}, we can see that at the end of 10 years when the paper is actually awarded, h-index is best performing index (as discussed in Section 2.3.2). During the initial years (less than 5 years) of the paper, number of citations is the best index. And h-index is the best index in the later years of the paper. This is expected as the number of citations will increase first and then the impact on h-index will be observed. H-index captures a deeper level of impact and hence it needs some amount of time to start seeing an increase.

\begin{figure}[H]
    \centering
    \includegraphics[scale=0.6]{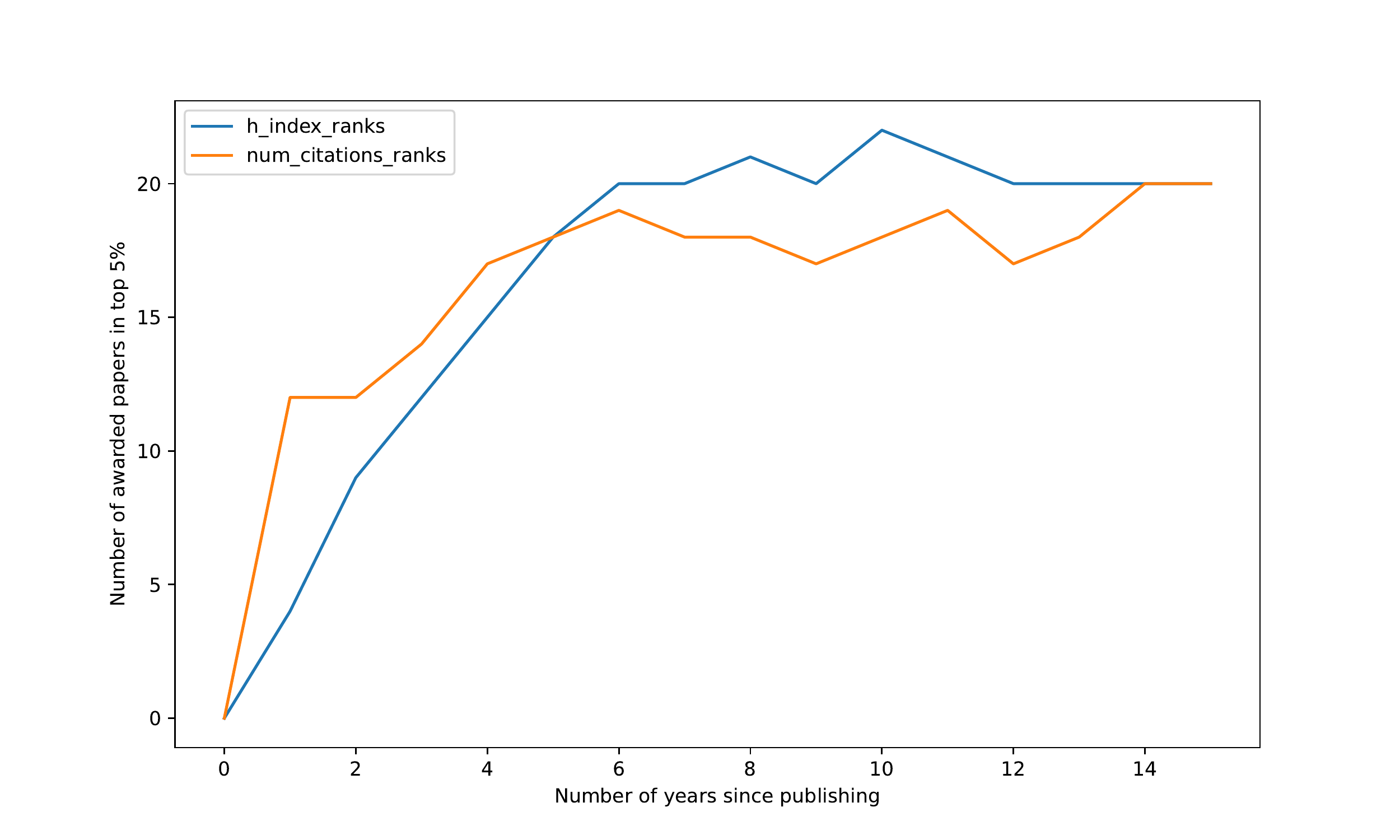}
    \caption{Number of years since publishing(beyond 10 years) vs Number of papers in top $5\%$ when ranked on the particular index}
    \label{fig:time_rank_15}
\end{figure}

Fig \ref{fig:time_rank_15} shows the performance of h-index and number of citations beyond 10 years of publishing. We observe that the performance of number of citations start to improve after the 10th year when the papers are awarded publicly. The hypothesis here is that once the papers are awarded they become more popular and it reaches more people. Consequently, it gets more number of citations. Therefore, more awarded papers start to rank in top 5\% of the list of papers according to number of citations. 

\subsection{Discussion}
In this section we define and compare h-index with different citation based indexes on research papers. Specifically, we looked at papers from SIGMOD and VLDB conference. We compare the rankings given by these indexes to each paper and also compare the rankings given to awarded papers. Our observations show that h-index is the best performing in ranking awarded papers at the top. That is, VLDB and SIGMOD lay higher recognition to h-index in determining the awards. In the next section, we will use the h-index of a paper to come up with the h-index of an author.

\section{h-index of authors using h-index of papers}

In this section, we discuss four modifications of the traditional h-index. First is the traditional h-index of a researcher. Secondly, we use h-frac-index proposed in \cite{koltun2021h}. They argue that h-index is not a good measure for scientific impact. This is due to changing authorship patterns, including a higher prevalence of hyper authorship. The major finding is that fractional allocation of citations among co-authors can mitigate the issues with h-index.

In section 2, we argue that the h-index of a paper is a better quantifier of paper impact when compared to many other metrics. Hence for the third metric, we use the h-index of a paper to calculate the h-index of an author. This metric was proposed in \cite{egghe2011single}. Finally, we propose hp-frac-index, as the h-index of an author using h-index of a paper divided by total authors of the paper (details in next section). 

We retrieve the top 1000 researchers ranked on decreasing h-index in three different fields of research, namely, Computer Science, Economics, and Biology. Then we cross reference our dataset with the scientific award winners in each field. The award lists used are Turing award winners for Computer Science, Nobel Prize in Economics, and Nobel prize in Chemistry, Physiology and Medicine for biology.  The traditional and proposed metrics are calculated for each researcher. Our experiments show that hp-index and hp-frac-index outperform traditional indices by giving better ranks to the awarded researchers in all three fields. We also compute the correlation amongst the metrics across the three fields.

\subsection{Metrics}

In this section, we will cover all the metrics being used in our experiments. We are considering four metrics. 
\subsubsection{h-index and h-frac-index}
\noindent \textbf{h-index}\\
We use the h-index as the first traditional metric for quantifying an author's research impact. H-index of an author is the largest number $h$ such that the given author has published at least $h$ papers that have each been cited at least $h$ times. 

\noindent \textbf{h-frac-index}\\
Secondly, we use a recent extension to h-index called h-frac \cite{koltun2021h}. This is a variant of the h-index that allocates citations fractionally among co-authors. In other words, when using the number of citations to calculate the h-index of an author, they divide each paper's number of citations by its total number of authors. This mitigates the cluttering of the ranking by hyper authors.

\subsubsection{hp-index and hp-frac-index}

\noindent We formally define the h-index of a \textit{paper}. Then we discuss the two metrics being used to evaluate researchers, namely, hp-index and hp-frac-index. 

Consider a paper $p$ and the set of papers citing $p$ be the set $C = [c_1, c_2, c_3,..., c_n]$. The h-index of $p$ is equal to the largest number $h$ such that at least $h$ papers from $C$ have at least $h$ citations each.

\noindent \textbf{hp-index}\\
In this metric, we first calculate the h-index of all papers of an \textit{author} $X$ using the definition above. Suppose, the set of papers published by $X$ is $[P_1, P_2, P_3,..., P_n]$ and the corresponding h-index values of these papers be $[h_1, h_2, h_3,..., h_n]$. We compute the hp-index of the author $X$ as follows:

\begin{equation}
    hp(X) = H([h_1, h_2, h_3,..., h_n])
\end{equation}

where $H$ is the function to calculate the h-index of any given set of values. To sum up, the hp-index of an author is the h-index of the h-index of all the author's papers. \\

\noindent \textbf{hp-frac-index}\\
Similar to hp-index, we calculate the h-index of all the papers $[P_1, P_2, P_3,..., P_n]$ as $[h_1, h_2, h_3,..., h_n]$. Let the number of authors for each paper be $[a_1, a_2, a_3,..., a_n]$. Note that for all $i, a_i \geq 1$. We then compute the hp-frac-index of an author $X$ as follows:

\begin{equation}
    hp-{frac}(X) = H([\frac{h_1}{a_1}, \frac{h_2}{a_2}, \frac{h_3}{a_3},..., \frac{h_n}{a_n}])
\end{equation}

In the hp-frac-index, the h-index of a paper is divided by the number of authors.

\begin{figure}[h]
    \centering
    \includegraphics[scale=0.6]{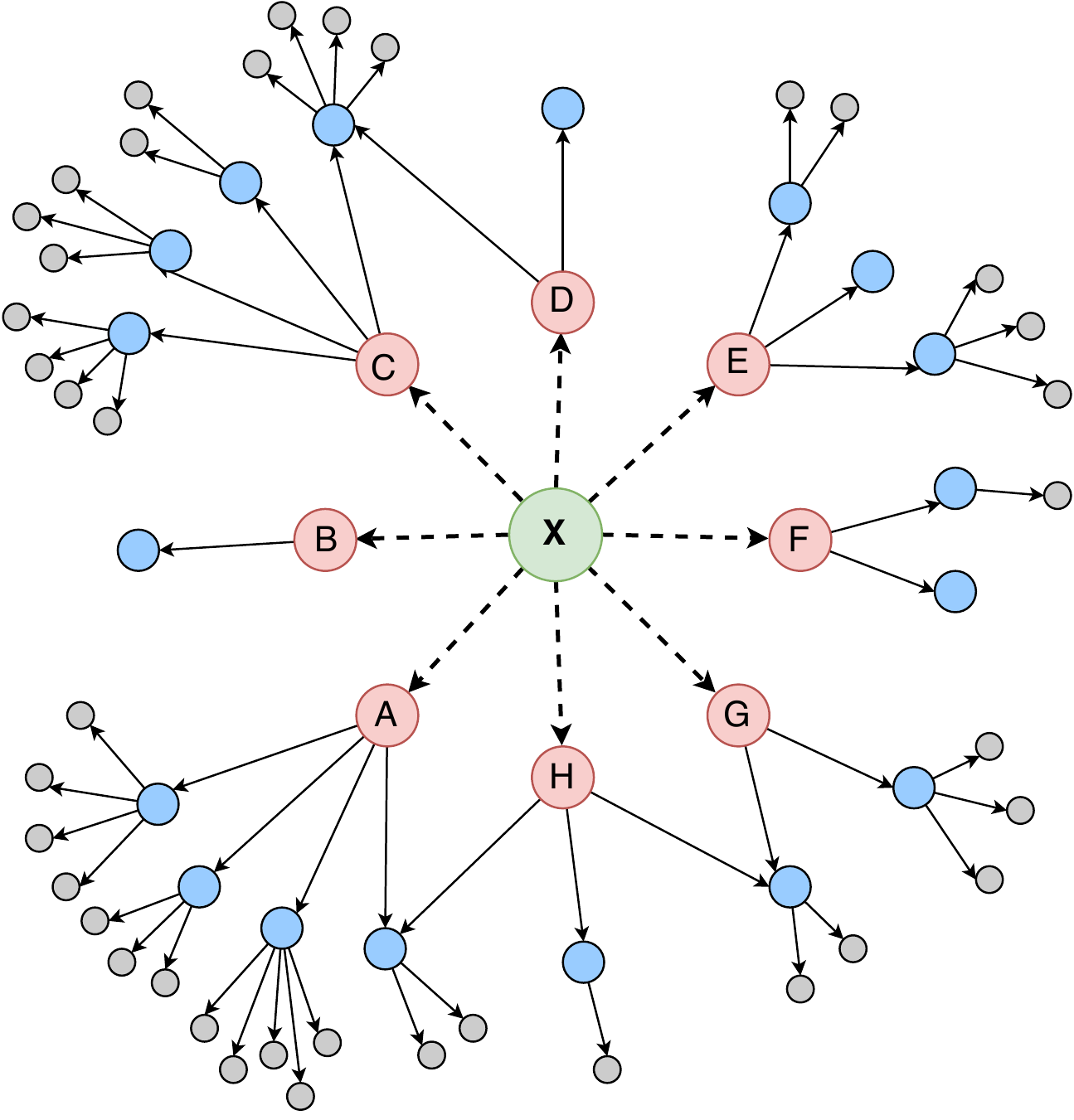}
    \caption{Example graph for h, hp, h-frac, hp-frac demonstration}
    \label{fig:h_frac_eg}
\end{figure}
\subsection{Example}
Consider the graph in Fig. \ref{fig:h_frac_eg}, node $X$ (in green) is the author, $A$ to $H$ are the papers co-authored by $X$. A dotted edge from $X$ to $A$ denotes that $X$ has co-authored the paper $A$.  All the blue papers are the ones that cite the red papers and grey papers cite the blue ones. $A$ solid edge from $A$ to $B$ denotes that $A$ is cited by $B$. We can see that $X$ has written 8 papers and paper $A$ has been cited 4 by 4 different papers. 

\begin{table}[h]
    \centering
    \begin{tabular}{l l l l l}
    \hline
        Node & \# authors & \# citations & list of citation of citation & h-index  \\
    \hline
        A & 2 & 4 & [4, 3, 5, 2] & 3 \\
        B & 1 & 1 & [0] & 0 \\
        C & 1 & 4 & [4, 3, 2, 4] & 3 \\
        D & 2 & 2 & [0, 4] & 1 \\
        E & 2 & 3 & [0,2,3] & 2 \\
        F & 1 & 2 & [1, 0] & 1 \\
        G & 3 & 2 & [2, 3] & 2 \\
        H & 2 & 3 & [1,2,2] & 2 \\
    \hline
    \end{tabular}
    \caption{H-index of the papers in given example in Fig. \ref{fig:h_frac_eg}}
    \label{tab:h_frac_eg}
\end{table}

In the table \ref{tab:h_frac_eg} shown above, $\# authors$ denotes the number of authors of a paper, $\# citations$ denotes the number of citations and \textit{list of citation of citation} denotes the number of citations of each blue paper that has been cited by the nodes $A$ to $H$. Lastly, $h-index$ is the h-index of each paper. Using the values of h-index for the nodes $A$ to $H$ from the table above, values of the four indices (defined in section 7.2.1 and 7.2.2) for the author can be calculated as follows:

\begin{align*} 
 h-index(X) &= H(4, 4, 3, 3, 2, 2, 2, 1) = 3\\ \\
h-frac-index(X) &= H(\frac{4}{2}, \frac{4}{1}, \frac{3}{2}, \frac{3}{2}, \frac{2}{2}, \frac{2}{1}, \frac{2}{3}, \frac{1}{1}) = 2\\ \\ 
hp-index(X) &= H(3, 3, 2, 2, 2, 1, 1, 0) = 2\\ \\ 
hp-frac-index(X) &= H(\frac{3}{2}, \frac{3}{1}, \frac{2}{2}, \frac{2}{2}, \frac{1}{2}, \frac{1}{1}, \frac{2}{3}, \frac{0}{1}) = 1\\
\end{align*}
Note that whenever there is a division of numbers, we take the floor (rounded down) value. For example, $3/2$ is considered as $1$.

\subsection{Citation Data}
In order to compare the four indices, we crawled the list of top 1000 researchers in Computer science, Economics and Biology field in the order of decreasing number of citations from google scholar \cite{scholar}. The steps followed to complete the data collection are:

\begin{itemize}
    \item Crawl the list of names of top 1000 authors from google scholar for each field.
    \item Match these author names to $author\_ids$ in Semantic Scholar \cite{s2ag}. We use Semantic Scholar as it has a highly accessible database of scientific literature with author and paper details readily available.
    \item Once we have $author\_ids$, we retrieve the papers published by them in one set of API calls. Let this set of papers for a researcher be called $P_w$.
    \item Next, we run another set of API calls to get the papers citing any paper in $P_w$. Let us call this set of papers as $P_c^1$.
    \item Lastly, we retrieve the papers citing any paper in $P_c^1$. Let us call this set of paper $P_c^2$.
\end{itemize}
At the end of this process, we have a graph for each author. One graph consists of the author and all the other papers from the sets $P_w, P_c^1, P_c^2$ as nodes. The author connects to the nodes in $P_w$ with a 'written by' edge. The nodes from $P_w$ connect to $P_c^1$ and $P_c^1$ connect to $P_c^2$ with a 'cited by' edge. This graph looks similar to the example in Fig. \ref{fig:h_frac_eg}

\begin{table}[H]
    \centering
    \begin{tabular}{l c c c}
    \hline
         & CS & Economics & Biology   \\
    \hline
    Crawled authors & 1000 & 1000 & 1000  \\
    Matched authors & 803 & 856 & 842 \\
    Total papers published ($P_w$) & 285622 & 159445 & 468683 \\ 
    Total Citations ($P_c^1$) & 6232959 & 3251169   & 12054908 \\
    Total Citations of citations ($P_c^2$) & 21053913 & 11497621  & 43194158 \\
    \hline
    \end{tabular}
    \caption{Dataset description}
    \label{tab:author_awards}
\end{table}

From the table above, we can see that Biology has the most number of papers published per author followed by Computer Science and then Economics. Subsequently, the two sets of citations follow the same trend. All three fields have more than 800 authors matching with their corresponding IDs in semantic scholar dataset. The total citation to total papers published ratio ($P_c^1$:$P_w$) is around 20 for all the fields. Whereas, the second level of citations to first level of citations ($P_c^2$:$P_c^1$) is around 3.5 for the three datasets.

\subsection{Experiments and results}

This section explains all the experiments done in order to compare the proposed indices with the traditional ones. We took the data collected, and calculated the h-index of each \textit{paper} published (i.e. all papers from set $P_w$). We use the definition discussed in Section 2.2 for the same. Using these values we calculated the four indices for an \textit{author} as explained in Section 2. Lastly, we ranked the authors on the decreasing order of each index (h, hp, h-frac, hp-frac) resulting in four different ranked lists of authors. Note that, if the value of an index is same for two different authors, then the one with more citations is given a higher rank.

\begin{figure}[htp]
\centering
\begin{tabular}{c}
  \includegraphics[scale=0.3]{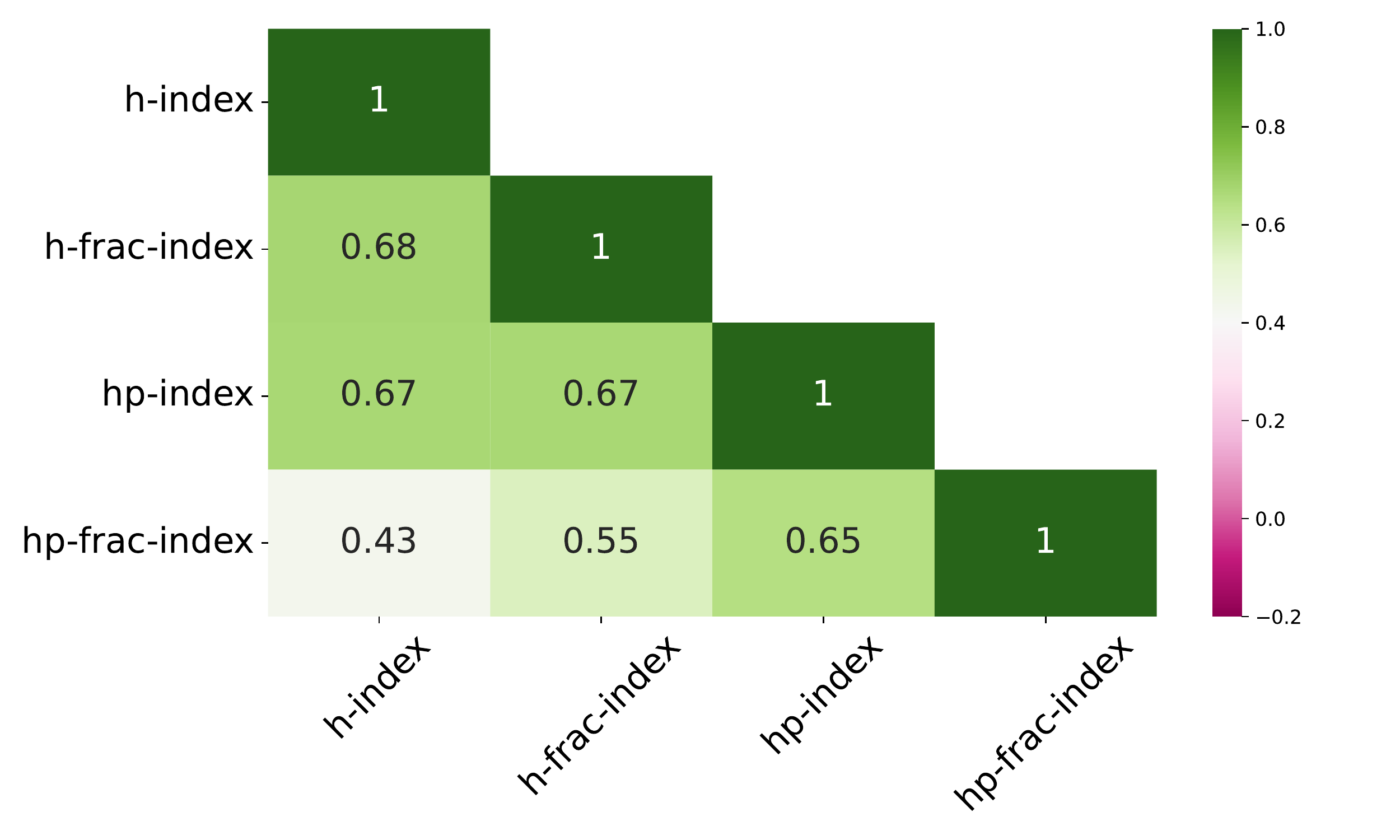} \\
 (a) RBO for Computer Science \\[6pt]
   \includegraphics[scale=0.3]{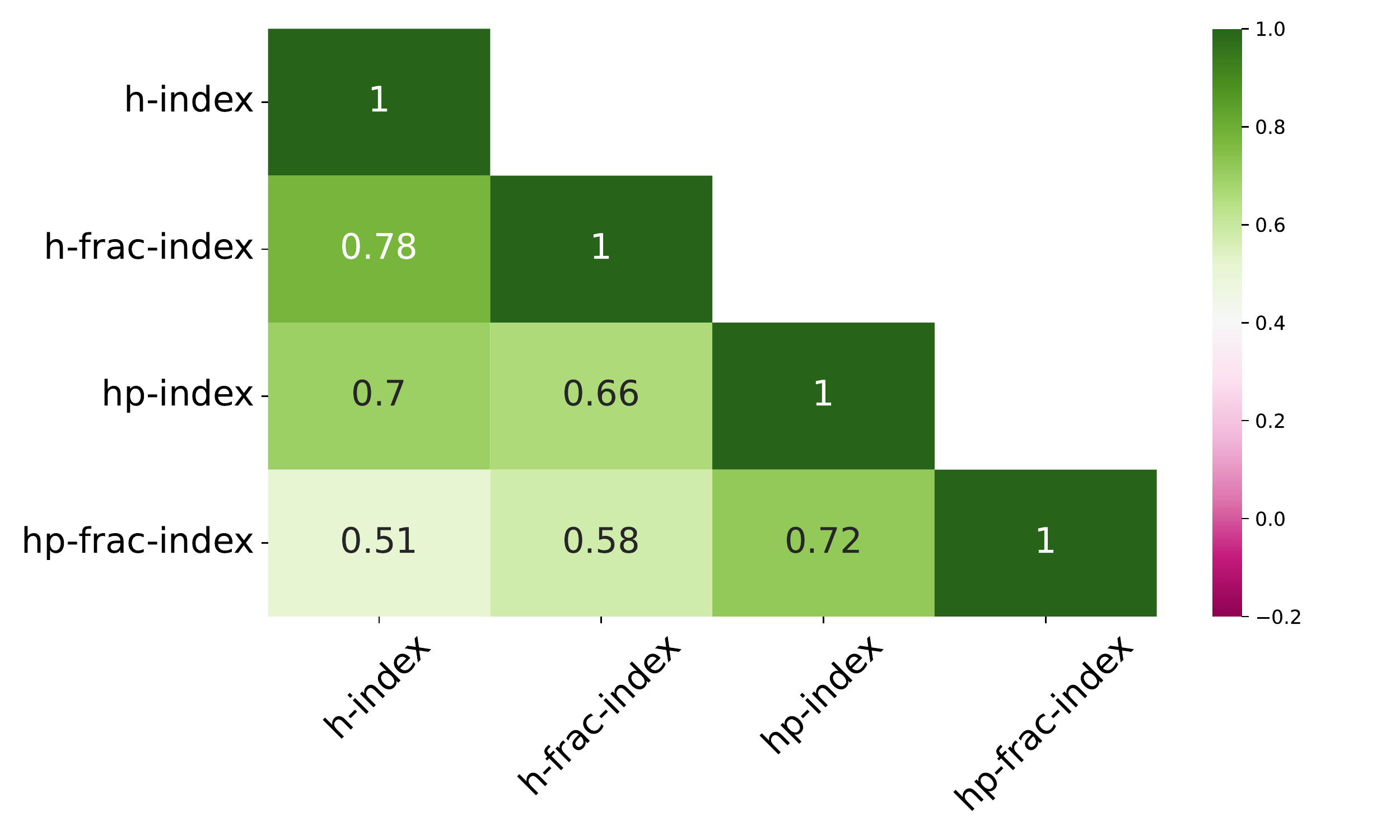} \\
 (b) RBO for Economics \\[6pt]
   \includegraphics[scale=0.3]{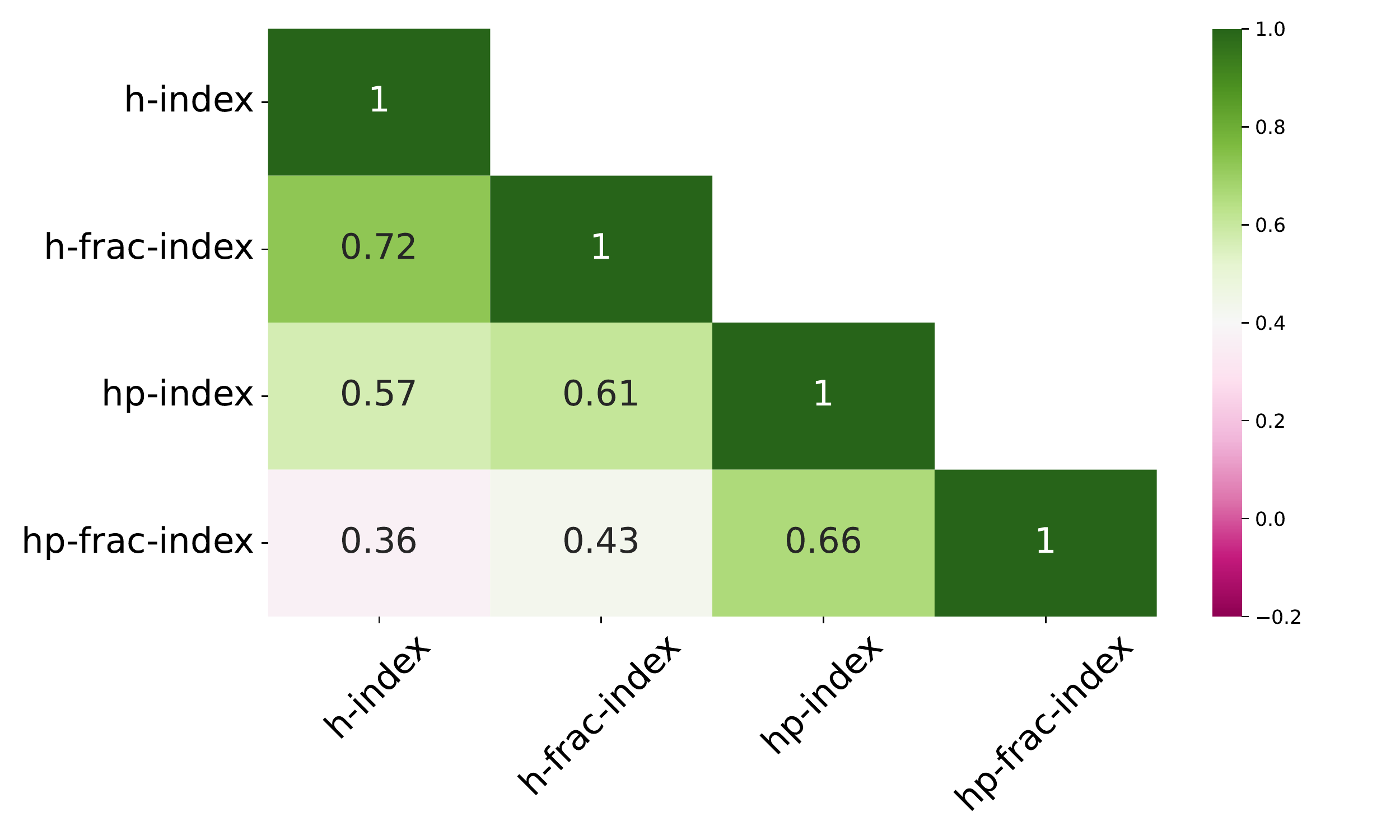} \\
(c) RBO for Biology \\[6pt]
\end{tabular}

\caption{Rank Biased Overlap for each of the fields}
\label{fig:correl_author}
\end{figure}

\subsubsection{RBO}
In this section, we compare the pairwise overlap of the four ranked lists. We use Rank Biased Overlap as proposed by Webber et al. \cite{webber2010similarity}. RBO acts as a similarity measure for ranked lists. From Fig \ref{fig:correl_author}, the overlap between h-index and hp-frac-index is the lowest with the values of 0.43, 0.51 and 0.36. The overlap is the highest for h-index and h-frac-index. The h-index, hp-index and hp-frac-index have high correlation (greater than $0.65$) amongst each of them. Whereas, the hp-frac-index has low overlap of $0.43, 0.55, 0.65$ with h-index, h-frac-index and hp-index respectively. Hence, the hp-frac-index is ranking differently from the other indices. This stimulates our next experiment to evaluate how each index ranks the awarded researchers and \textit{which one can rank awarded researchers higher}.

\begin{table*}[h]
\setlength{\tabcolsep}{3pt}
    \centering
    \begin{tabular}{l |c c c c | c c c c | c c c c}
    \hline
         & \multicolumn{4}{c|}{Biology} & \multicolumn{4}{c|}{Computer Science} & \multicolumn{4}{c}{Economics}\\
    \hline
    & h & h-frac & hp & hp-frac & h & h-frac & hp & hp-frac & h & h-frac & hp & hp-frac \\
    \hline
    Top 5\% &      23.08 & 38.4 & 38.4 & \textbf{61.5} & 27.7 & 38.8 & 27.7 & \textbf{50} & 26.4 &  29.4 & 29.4 & \textbf{44.1}\\
    5\% - 10\% &   \textbf{23.1} & 15.3 & 15.3 & 7.6 & \textbf{16.6} & 5.5 & \textbf{16.6} & \textbf{16.6} & 20.5 & \textbf{26.4} & 17.6 & 20.5 \\
    10\% - 15\% &  \textbf{15.3} & 7.6 & 7.6 & \textbf{15.3} & \textbf{16.6} & 11.1 & \textbf{16.6} & 11.1 & 11.7 & \textbf{20.5} & 11.7 & 8.8 \\
    15\% - 20\% &  \textbf{7.6} & \textbf{7.6} & \textbf{7.6} & 0 & \textbf{11.1} & 5.5 & 0 & 5.5 & \textbf{14.7} & 5.8 & 8.8 & 8.8 \\
    \hline
    Total $\le$ 20\% & 69.08 & 68.9   & 68.9 & \textbf{84.4} & 72 & 60.9 & 60.9 & \textbf{77.7} & 73.3 & \textbf{82.1} & 67.5 & \textbf{82.2} \\
    \hline
    \end{tabular}
    \caption{Percentage of awardees in different ranges of ranked lists as per each index across three fields}
    \label{tab:top_each}
    \vspace{-2mm}
\end{table*}

\subsubsection{Awarded researchers}
In this experiment, we compare the position of each award winning researcher in the four different ranked lists obtained for each of the three fields. We compiled the list of awardees for the awards listed in Table \ref{tab:awardlist} and cross referenced them to our list of top 1000 researchers in each field. We found eighteen such awarded researchers in Computer Science, thirty four in Economics and thirteen in Biology. Then, we extracted the values and ranks as per the indices for all awardees (see Table \ref{tab:author_awards_bio}, \ref{tab:author_awards_cs}, \ref{tab:author_awards_eco}).

Table \ref{tab:top_each} shows the percentage of awardees ranked amongst top 5\%, 5\%-10\%, 10\%-15\% and 15\%-20\% for each index across the three fields. We can observe that hp-frac-index is the best performing index with  around $61\%$ (Biology), $50\%$ (Computer Science) and $44\%$ (Economics) of awardees being ranked in the top $5\%$ of the list. The other three indices perform 10\% poorer than hp-frac-index in ranking awarded researchers in top 5\%. For Biology, h-index performs the best in all ranges except top 5\%. Although other indices perform equally in some ranges. We observe similar trend in Computer Science as well. However, in Economics, the next best indicator is h-frac in 5\%-10\% and 10\%-15\% range. Overall, in the range of $<20\%$, we see that hp-frac-index performs better than the other indices for Biology and Computer Science. It performs at par with h-frac index for Economics.

Table \ref{tab:author_awards_bio}, \ref{tab:author_awards_cs}, and \ref{tab:author_awards_eco} show the ranks and values as per each index for all the awarded researchers in our data set. In Table \ref{tab:best_ranks}, we show the percentage of awardees (for a particular field) that receive the best or highest rank as per a given index. We do this for all the three fields and all the four indices. To elaborate, consider an awardee $R$, they have four ranks as per each index. Suppose, out of these four ranks the rank given by h-index is the highest, then increment the count for h-index by 1. For example in Table \ref{tab:best_ranks}, under Biology, 15.3\% for h-index means that 15.3\% of all Biology awardees had the best rank as per h-index (among the four indices). We can observe that hp-frac-index far outperforms the other indices across all three fields. Note, that if more than one index for an awardee yields equal best rank, it is considered towards all those indices, hence the sum of each column may surpass 100\%.

The outcome here is that the hp-frac-index perform better than all the other methods. The hp-index and hp-frac-index capture a deeper level of impact by taking into account one extra level of papers as compared to h-index and h-frac-index. This extra depth of information helps in recognising the sustained research impact of the author better than h-index.

\begin{table}[h]
    \centering
    \begin{tabular}{l c c c}
    \hline
         & Biology & Computer Science & Economics \\
         \hline
        h-index       & 15.3 & 16.6 & 8.8 \\
        h-frac-index  & 0 & 0 & 29.4 \\
        hp-index      & 30.7 & 22.2 & 17.6 \\
        hp-frac-index & \textbf{61.5} & \textbf{66.6} & \textbf{47} \\
        \hline
    \end{tabular}
    \caption{Percentage of awardees given the highest rank as per each index across the three fields (see Table \ref{tab:author_awards_bio}, \ref{tab:author_awards_cs}, and \ref{tab:author_awards_eco}}
    \label{tab:best_ranks}
\end{table}

\subsubsection{Further Analysis}
\begin{table}[h]
    \centering
    \begin{tabular}{lc c c}
    \hline
         & CS & Economics & Biology   \\
    \hline
    Average h-index                    & 70.3 & 47.8 & 111.7  \\
    Max h-index                        & 184 & 149   & 285 \\
    \hline
    Average h-frac-index               & 41.6 & 34.6 & 44.5 \\
    Max h-frac-index                   & 107 & 114   & 148 \\
    \hline
    Average hp-index                   & 32.9 & 26.2 & 53.09 \\ 
    Max hp-index                       & 72 & 69   & 122 \\
    hp-index of 100th author           & 30 & 32   & 53 \\
    \hline
    Average hp-frac-index              & 16.7 & 16.9   & 18.08 \\
    Max hp-frac-index                  & 42 & 47   & 51 \\
    hp-frac of 100th author            & 23 & 24   & 29 \\
    \hline
    Average number of publications     & 383.9 & 201   & 622.09 \\ 
    \hline
    \end{tabular}
    \caption{Average and maximum values of the calculated metrics}
    \label{tab:author_stats}
\end{table}

The table \ref{tab:author_stats}, displays a number of statistics about the calculated indices and their average and maximum values. We can see that the average h-index for Biology is the highest with a big margin. The average number of publications is the highest for Biology. This also explains the high maximum h-index value for the same.  

The average hp-frac-index does not have a big margin for Biology. This shows that the range of hp-frac values are more tightly packed. The 100th author as per hp-frac index has a hp-frac value of 23 for Computer Science, 24 for Economics and 29 for Biology. For hp-index, we can observe that the average value and the value for 100th author are very close. This means that there is sharp fall in the values of hp-index from 1st to 100th author. Whereas for hp-frac-index, therer is a consistent difference between average value and 100th author value. This shows that hp-frac-index has slower decline in the values (when we move down the rank list) as compared to hp-index.

Both hp-frac-index and h-frac-index include a division by number of authors in their calculation. One might wonder if these two indices only punish authors who have high number of co-authors? To evaluate this, we calculated the difference of h-index and h-frac-index, and hp-index and hp-frac-index for each author. Let us call them $\textit{diff1}$ and $\textit{diff2}$ respectively. We ranked the authors on descending order of $\textit{diff1}$ and $\textit{diff2}$. Then we ranked the authors on decreasing order of average co-authors. Finally, we calculate the Spearman's correlation of $\textit{diff1}$ and $\textit{diff2}$ ranked lists with the ranked list ordered by average co-authors. Note that we did this for all three fields. 
As shown in table \ref{tab:diff_corel}, the average number of co-authors has a very low correlation with the difference of h-index and h-frac-index, and hp-index and hp-frac-index values. This shows that hp-frac-index and h-frac-index are not antithetical to collaboration with others. The fact that they have very low correlation shows that these two rank lists do not follow the same trend.

\begin{table}[H]
    \centering
    \begin{tabular}{lccc}
    \hline
         & CS & Economics & Biology   \\
    \hline
    Correlation between $diff1$ and average co-authors                    & 0.08 & 0.15 & 0.07  \\
    Correlation between $diff2$ and average co-authors                    & 0.17 & 0.22 & 0.039  \\
    \hline
    \end{tabular}
    \caption{Correlation between $diff1$, $diff2$, and the average number of co-authors }
    \label{tab:diff_corel}
\end{table}

Furthermore, authors like Gregg L. Semenza with a average co-authorship of $14.17$ has the highest rank in Biology. We also notice that highly collaborative authors like Yoshua Bengio and Michael I. Jordan rank amongst the top 5 authors for Computer Science. 

\section{Conclusion}
In this work, we collected large-scale data for evaluation of author
from three fields, namely, Computere Science, Economics and Biology. We used four different metrics: two traditional metrics, one
re-applied metric (hp) and one proposed metric (hp-frac). Our experimental analysis show that hp-frac-index gives a unique ranking
order to authors and outperforms all the the other metrics in ranking the awarded researchers higher. The hp-frac-index is a robust
way to evaluate the impact of researchers. Its ability to capture individual contributions and resist manipulation makes it a valuable
tool for assessing the impact of researcher. It takes into account
the importance of a paper’s impact by using the paper’s h-index,
therefore, capturing a second level of research impact. These factors
together make hp-frac-index better at ranking the authors. One of the problems to address in further work is the ability to predict
future award winners using these metrics more accurately.

\clearpage

\begin{table}[H]
    \centering
    \begin{tabular}{l c c c}
    \hline
        Award & Total awardees & Matched awardees & Acronym   \\
    \hline
     Turing award winners (for Computer Science) & 70 & 8 & CS1\\
     ACM Prize in Computing & 13 & 10 & CS2\\
     \hline
     Nobel Prize in Economics & 84 & 15& EC1\\
    Fellows of the American Finance Association & 66 & 19 &EC2 \\
     \hline
     Nobel Prize in Chemistry & 184 & 2 & B1\\
     Nobel Prize in Physiology or Medicine & 219 & 2 & B2\\
     Breakthrough Prize in Life Sciences & 48 & 9 & B3\\
    \hline
    \end{tabular}
    \caption{List of awards collected}
    \label{tab:awardlist}
    \vspace{-5mm}
\end{table}

\begin{table}[H]
    \centering
    \begin{tabular}{|l |c |c | l l | l l| l l| l l |}
    \hline
       \multirow{2}{*}{Author name}  & \multirow{2}{*}{Award} & \multirow{2}{*}{Avg. co-authors}  & \multicolumn{2}{c|}{h-index} & \multicolumn{2}{c|}{h-frac-index} & \multicolumn{2}{c|}{hp-index} & \multicolumn{2}{c|}{hp-frac-index}\\
    \cline{4-11}
    & & & value & rank & value & rank & value & rank & value & rank \\
    \hline
Gregg L. Semenza &B2 &  14.18 &  177 &  44  &  103 &  10  &  84 &  32  &  \textbf{51} &  \textbf{1}  \\
Robert A. Weinberg &B3 &  \textbf{5.2} &  177 &  43  &  101 &  13  &  92 &  15  &  \textbf{41} &  \textbf{9}  \\
Lewis C. Cantley &B3 &  9.25 &  175 &  49  &  77 &  44  &  75 &  53  &  \textbf{40} &  \textbf{14}  \\
David Botstein &B3 &  7.53 &  159 &  86  &  76 &  47  &  77 &  46  &  \textbf{38} &  \textbf{21}  \\
Eric S. Lander &B3  &  \textbf{26.54} &  \textbf{285} &  \textbf{1}  &  95 &  19  &  \textbf{122} &  \textbf{1}  &  37 &  25  \\
Bert Vogelstein &B3  &  10.98 &  255 &  5  &  103 &  8  &  \textbf{111} & \textbf{ 3 } &  36 &  32  \\

Robert J. Lefkowitz &B1 &  5.48 &  \textbf{214} &  \textbf{16}  &  91 &  26  &  85 &  28  &  36 &  33  \\

James P. Allison &B2, B3  &  13.85 &  140 &  151  &  63 &  125  &  66 &  136  &  \textbf{35} &  \textbf{40}  \\
Gary B. Ruvkun &B3 &  5.42 &  101 &  500  &  55 &  214  &  59 &  260  &  \textbf{32} &  \textbf{60}  \\

Karl Deisseroth &B3 &  9.54 &  158 &  89  &  62 &  134  &  \textbf{71} &  \textbf{89}  &  29 &  94  \\
Aaron Ciechanover &B1 &  6.92 &  106 &  447  &  58 &  180  &  55 &  355  &  \textbf{28} &  \textbf{117}  \\
Xiaowei Zhuang &B3 &  9.63 &  89 &  618  &  47 &  340  &  47 &  546  &  \textbf{20} &  \textbf{321}  \\
Masashi Yanagisawa &B3 &  9.12 &  129 &  203  &  50 &  297  &  \textbf{63} &  \textbf{181}  &  17 &  418  \\

    \hline
    \end{tabular}
    \caption{List of award winners with ranks for Biology (highest ranks in bold)}
    \label{tab:author_awards_bio}
\end{table}

\begin{table}[H]
    \centering
    \begin{tabular}{|c c |c c |c c |}
    \hline
       \multicolumn{2}{|c|}{Biology} & \multicolumn{2}{c|}{Computer Science} & \multicolumn{2}{c|}{Economics}\\
    \hline
    Author & Awarded? & Author  & Awarded? & Author & Awarded? \\
    \hline
    \textbf{Gregg L. Semenza}  & \textbf{Yes} & \textbf{Geoffrey E. Hinton}  & \textbf{Yes} & Cass R. Sunstein  & No\\
Michael Karin  & No & Ronald R. Yager  & No & \textbf{James J. Heckman} &  \textbf{Yes}\\
Edmund T. Rolls &  No & \textbf{Judea Pearl}  &\textbf{Yes} & \textbf{Richard H. Thaler} & \textbf{Yes}\\
Joan Massagué &  No & \textbf{Yoshua Bengio} &  \textbf{Yes} & Dani Rodrik & No\\
K. J. Friston  & No & Andrew P. Zisserman  & No & \textbf{William D. Nordhaus}  & \textbf{Yes}\\

Douglas G. Altman & No & Michael I. Jordan & No &  Colin F. Camerer & No \\
Joseph E. LeDoux & No &  \textbf{Yann Le Lecun} & \textbf{Yes} &  \textbf{Paul A. Samuelson} & \textbf{Yes}\\
Solomon H. Snyder & No &  Tomaso A. Poggio & No &  Gary S. Becker & No \\
\textbf{Robert A. Weinberg} & \textbf{Yes} &  Lotfi A. Zadeh & No & Robert W. McGee & No \\ 
Mark P. Mattson & No &  \textbf{Jon M. Kleinberg} & \textbf{Yes} &  \textbf{Jean Tirole} & \textbf{Yes} \\
\hline
\% of awardees & 20\% & \% of awardees  & 50\% &  \% of awardees & 50\% \\

    \hline
    \end{tabular}
    \caption{List of top 10 authors ranked by hp-frac-index (awarded researchers are in bold)}
    \label{tab:top10_hp_frac}
\end{table}

 \begin{table}[H]
    \centering
    \begin{tabular}{|l |c|c | l l | l l| l l| l l |}
    \hline
       \multirow{2}{*}{Author name} & \multirow{2}{*}{Awards}& \multirow{2}{*}{Average co-authors}  & \multicolumn{2}{c|}{h-index} & \multicolumn{2}{c|}{h-frac-index} & \multicolumn{2}{c|}{hp-index} & \multicolumn{2}{c|}{hp-frac-index}\\
    \cline{4-11}
    & && value & rank & value & rank & value & rank & value & rank \\
    \hline
Geoffrey E. Hinton &CS1  &  3.14 &  142 &  9  &  106 &  2  &  65 &  3  &  \textbf{42} &  \textbf{1}  \\
Judea Pearl &CS1 &  \textbf{1.68} &  104 &  69  &  89 &  8  &  42 &  89  &  \textbf{37} &  \textbf{3}  \\
Yoshua Bengio &CS1 &  5.03 &  \textbf{184} &  \textbf{1}  &  105 &  4  &  \textbf{72} &  \textbf{1}  &  34 &  4  \\

Yann Le Lecun &CS1 &  6.74 &  117 &  40  &  73 &  27  &  \textbf{62} &  \textbf{4}  &  33 &  7  \\
Jon M. Kleinberg  &CS2&  3.64 &  108 &  61  &  72 &  30  &  49 &  27  &  \textbf{30} &  \textbf{10}  \\
Daphne L. Koller &CS2 &  7.07 &  129 &  23  &  68 &  36  &  \textbf{52} &  \textbf{19}  &  26 &  28  \\

Jeffrey David Ullman &CS1 &  3.51 &  99 &  89  &  66 &  44  &  46 &  42  &  \textbf{26} &  \textbf{26}  \\
Dan Boneh &CS2 &  5.44 &  117 &  42  &  72 &  31  &  46 &  44  &  \textbf{26} &  \textbf{29}  \\
Ronald L. Rivest &CS1  &  4.35 &  79 &  234  &  50 &  163  &  37 &  193  &  \textbf{26} &  \textbf{25}  \\
Pat M. Hanrahan &CS1 &  4.58 &  86 &  160  &  52 &  131  &  41 &  115  &  \textbf{25} &  \textbf{54}  \\
Stefan Savage &CS2  &  5.41 &  87 &  152  &  43 &  327  &  41 &  118  &  \textbf{24} &  \textbf{78}  \\
David M. Blei &CS2 &  3.48 &  92 &  115  &  56 &  92  &  38 &  165  &  \textbf{24} &  \textbf{65}  \\
M. Frans Kaashoek &CS2 &  4.09 &  77 &  254  &  42 &  350  &  38 &  167  &  \textbf{23} &  \textbf{95}  \\
Pieter Abbeel &CS2 &  5.59 &  \textbf{129} &  \textbf{24}  &  56 &  97  &  45 &  54  &  22 &  119  \\

David A. Patterson &CS1 &  4.58 &  \textbf{92} &  \textbf{114}  &  50 &  164  &  38 &  164  &  21 &  136  \\

John Leroy Hennessy &CS2 &  3.76 &  67 &  415  &  41 &  376  &  33 &  372  &  \textbf{17} &  \textbf{339}  \\

David Silver &CS2 &  \textbf{7.86} &  66 &  434  &  40 &  408  &  \textbf{35} &  \textbf{262}  &  15 &  462  \\

Jeffrey Dean &CS2 &  7.5 &  35 &  758  &  30 &  658  &  29 &  541  &  \textbf{14} &  \textbf{533}  \\
    \hline
    \end{tabular}
    \caption{List of award winners with ranks for Computer Science (highest ranks in bold)}
    \label{tab:author_awards_cs}
\end{table}

\begin{table}[H]
    \centering
    \begin{tabular}{|l | c|c | l l | l l| l l| l l |}
    \hline
       \multirow{2}{*}{Author name} & \multirow{2}{*}{Awards}  & \multirow{2}{*}{Avg. co-authors}  & \multicolumn{2}{c|}{h-index} & \multicolumn{2}{c|}{h-frac-index} & \multicolumn{2}{c|}{hp-index} & \multicolumn{2}{c|}{hp-frac-index}\\
    \cline{4-11}
    & & & value & rank & value & rank & value & rank & value & rank \\
    \hline
James J. Heckman  & EC1 &  2.91 &  \textbf{149} &  \textbf{1}  &  \textbf{114} &  \textbf{1}  &  \textbf{69} &  \textbf{1}  &  45 &  2  \\
Richard H. Thaler & EC1, EC2 &  3.15 &  89 &  23  &  73 &  16  &  53 &  10  &  \textbf{39} &  \textbf{3}  \\
William D. Nordhaus & EC1 &  2.27 &  86 &  30  &  76 &  12  &  43 &  28  &  \textbf{39} &  \textbf{5}  \\
Paul A. Samuelson &EC2 &  2.13 &  86 &  29  &  \textbf{84} &  \textbf{6}  &  39 &  50  &  38 &  7  \\
Jean Tirole & EC1, EC2 &  3.02 &  121 &  4  &  \textbf{98} &\textbf{  }2  &  58 &  4  &  36 &  10  \\
Jeremy C. Stein & EC2 &  3.91 &  74 &  66  &  61 &  46  &  46 &  18  &  \textbf{33} &  \textbf{17}  \\
Ben S. Bernanke & EC1, EC2 &  \textbf{1.38} &  68 &  94  &  64 &  32  &  40 &  42  &  \textbf{33} &  \textbf{16}  \\
Alvin E E. Roth & EC1 &  3.25 &  90 &  22  &  69 &  23  &  39 &  52  &  \textbf{33} &  \textbf{21}  \\
Christopher A. Sims & EC1  &  1.57 &  61 &  148  &  56 &  74  &  34 &  117  &  \textbf{33} &  \textbf{19}  \\
René M. Stulz & EC2 &  3.59 &  96 &  18  &  \textbf{76} &  \textbf{11}  &  47 &  14  &  32 &  23  \\
Raghuram G. Rajan & EC2 &  3.26 &  73 &  70  &  63 &  35  &  \textbf{46} &  \textbf{16}  &  32 &  22  \\
Joshua D. Angrist & EC1 &  4.21 &  75 &  64  &  56 &  72  &  \textbf{45} &  \textbf{23}  &  32 &  24  \\

John Y. Campbell & EC2  &  3.61 &  66 &  106  &  54 &  80  &  40 &  45  &  \textbf{31} &  \textbf{29}  \\

G. William Schwert & EC2 &  2.28 &  60 &  158  &  47 &  112  &  35 &  103  &  \textbf{31} &  \textbf{31}  \\
John H. Cochrane & EC2 &  3.38 &  52 &  250  &  49 &  102  &  30 &  200  &  \textbf{31} &  \textbf{33}  \\
Esther Duflo & EC1 &  \textbf{12.06} &  87 &  27  &  61 &  45  &  \textbf{45} &  \textbf{22}  &  27 &  57  \\

Luigi Zingales & EC2 &  3.02 &  71 &  80  &  54 &  79  &  \textbf{40} &  \textbf{43}  &  27 &  58  \\
Guido W. Imbens & EC1 &  3.68 &  \textbf{81} &  \textbf{43}  &  56 &  73  &  40 &  44  &  27 &  60  \\

Franklin Allen  & EC2&  2.86 &  80 &  51  &  60 &  50  &  \textbf{40} &  \textbf{49}  &  27 &  68  \\

Campbell R. Harvey & EC2 &  2.79 &  80 &  49  &  \textbf{63} &  \textbf{36}  &  35 &  99  &  27 &  62  \\
Abhijit V. Banerjee & EC1 &  4.36 &  \textbf{89} &  \textbf{24}  &  61 &  47  &  35 &  100  &  27 &  63  \\
Lars Peter Hansen & EC1, EC2 &  3.05 &  66 &  109  &  52 &  89  &  32 &  140  &  \textbf{27} &  \textbf{66}  \\

Lloyd S. Shapley & EC1 &  2.05 &  53 &  234  &  45 &  126  &  32 &  139  &  \textbf{24} &  \textbf{96}  \\
Eduardo S. Schwartz & EC2 &  2.89 &  66 &  110  &  \textbf{51} &  \textbf{91}  &  30 &  193  &  23 &  108  \\
John R. Graham & EC2 &  2.63 &  60 &  159  &  44 &  142  &  30 &  195  &  \textbf{23} &  \textbf{113}  \\
José A. Scheinkman & EC2 &  3.17 &  59 &  169  &  \textbf{46} &  \textbf{117}  &  32 &  146  &  21 &  162  \\
David Hirshleifer & EC2 &  3.44 &  64 &  130  &  \textbf{50} &  \textbf{97}  &  30 &  192  &  21 &  159  \\
Franco Modigliani & EC2 &  2.95 &  53 &  235  &  \textbf{43} &  \textbf{163}  &  29 &  229  &  21 &  166  \\
Robert B. Wilson & EC1 &  1.39 &  33 &  726  &  31 &  435  &  22 &  547  &  \textbf{21} &  \textbf{183}  \\
David S. Scharfstein & EC2 &  4.39 &  42 &  446  &  34 &  337  &  \textbf{30} &  \textbf{197}  &  19 &  230  \\

Laura T. Starks & EC2 &  3.1 &  55 &  212  &  \textbf{42} &  \textbf{176}  &  29 &  235  &  19 &  234  \\
William F. Sharpe  & EC2&  2.35 &  38 &  567  &  35 &  305  &  25 &  385  &  \textbf{18} &  \textbf{263}  \\
Robert H. Litzenberger & EC2 &  2.47 &  31 &  769  &  25 &  664  &  22 &  546  &  \textbf{16} &  \textbf{409}  \\

Philip H. Dybvig & EC1 &  4.31 &  30 &  786  &  25 &  666  &  18 &  775  &  \textbf{13} &  \textbf{614}  \\
    \hline
    \end{tabular}
    \caption{List of award winners with ranks for Economics (highest ranks in bold)}
    \label{tab:author_awards_eco}
\end{table}

\bibliographystyle{unsrt}  
\bibliography{references}

\end{document}